\documentclass[preprint,showpacs,preprintnumbers,amsmath,amssymb,prb]{revtex4-1}

\usepackage[dvipdfmx]{graphicx}
\usepackage{dcolumn}
\usepackage{bm}
\usepackage{color} 
\begin{document}
%
\title{Temperature scaling of effective polaron mobility    
in energetically disordered media}
\author{
Kazuhiko Seki
}
\email{k-seki@aist.go.jp}
\affiliation{Nanomaterials Research Institute(NMRI), 
National Institute of Advanced Industrial Science and Technology (AIST)\\
AIST Tsukuba Central 5, Higashi 1-1-1, Tsukuba, Ibaraki, Japan, 305-8565
}
\author{Mariusz Wojcik}
\affiliation{
Institute of Applied Radiation Chemistry, Lodz University of Technology,
Wroblewskiego 15, 93-590 Lodz, Poland
}
%
\begin{abstract}
We study 
effective mobility 
in 2 dimensional (2D) and 3 dimensional (3D) systems, 
where hopping transitions of carriers are described 
by the Marcus equation under a Gaussian density of states in the dilute limit. 
Using an effective medium approximation (EMA), we 
determined the coefficient $C_d$ for 
the effective mobility expressed by 
$\mu_{\rm eff}\propto\exp\left[-\lambda/\left(4 k_{\rm B} T\right)-
C_d\sigma^2/\left(k_{\rm B} T\right)^2
\right]/\left[\sqrt{\lambda} (k_{\rm B} T)^{3/2}\right]$, where 
$\lambda$ is the reorganization energy, $\sigma$ is the standard deviation of the Gaussian density of states, and 
$k_{\rm B} T$ takes its usual meaning. 
We found $C_d=1/2$ for both 2D and 3D. 
While various estimates of the coefficient $C_d$ for 3D systems are available in the literature, 
we provide for the first time the expected $C_d$ value for a 2D system. 
By means of kinetic Monte-Carlo simulations, 
we show that the effective mobility is well described by the equation shown above 
under certain conditions on $\lambda$. 
We also give examples of analysis of experimental data for 2D and 3D systems based on our theoretical results.

\end{abstract}

\maketitle
\section{Introduction}
Recently, 
organic semiconductors have been attracting considerable attention for their use in  
electronic devices like organic light-emitting diodes and organic solar cells. \cite{Brutting}
Carrier transport in molecular solids can be described by hopping transitions between neighboring molecules and 
the mobility is considered to be strongly influenced by electrostatic energy distribution on ionized molecules. \cite{Bassler_93} 
In amorphous molecular solids, 
the electrostatic energy at each molecule is different because 
the polarization originating from the surrounding molecules fluctuates if 
the molecular orientation and arrangement are distributed.  \cite{Dunlap_96,Novikov_94,Young_95,Novikov_98,Seki_01}
A Gaussian distribution of the site energy is expected 
from the central limit theorem and the variance $\sigma^2$ characterizes 
the site energy disorder. \cite{Bassler_93,Dunlap_96,Novikov_94,Young_95,Seki_01}
As a result of energetic disorder, 
the mobility deviates from the Arrhenius law and scales with the reciprocal square
of temperature.
In analyzing experiments and interpreting computer simulation results, 
the low-field drift mobility in disordered organic solids has been commonly expressed in the form \cite{Seki_01,Parris_01,Lukyanov_10,Fishchuk_13,Baranovskii14}
\begin{align}
\mu_{\rm eff} \propto \exp[-E_a/(k_{\rm B} T)-C_d \sigma^2 /(k_{\rm B}T)^2],
\label{eq:scaling1}
\end{align} 
with a parameter $E_a$ characterizing the activation energy. 
$C_d$ is a numerical constant independent of $\sigma$ and temperature $T$. 
$k_{\rm B}$ is the Boltzmann constant. 

The expression given by Eq. (\ref{eq:scaling1}) has been frequently used to determine $\sigma$ from experimental data 
by plotting $\ln \mu_{\rm eff}$ against $1/T^2$. \cite{Tessler14,Bassler_93,Baranovskii14,Ochse99,Bleyl99,Hertel_08}
In order to determine $\sigma$, the numerical value of $C_d$ should be known in advance and it is 
important to theoretically determine $C_d$ to extract the correct value of $\sigma$ from experimental data. 

By means of simulations which assumed the Gaussian density of states and  
a carrier transport model based on phonon assisted tunneling 
and hopping (Miller-Abrahams (MA) process), \cite{Ambegaokar_71,Bassler_93}
the numerical parameter $C_d$ was found to be equal to  
$0.44$ in 3 dimensions (3D). \cite{Bassler_93}
In 1 dimension (1D), 
$C_d=1$ with 
an extra weak $\sigma$-dependence is obtained for the same model by analytical exact calculation. \cite{Cordes_01}
Clearly,
the value of $C_d$ depends on the dimensionality and the coordination number. 

In principle, 
the parameter $C_d$ may be influenced by elementary transition rates. 
The carrier transport in organic solids can be regarded as series of self-exchange reactions \cite{Soos_00,Seki_01,Verbeek_92}
and the elementary transition rate of self-exchange reaction in solution is expressed by 
the Marcus equation. 
\cite{Marcus_56,Marcus_64}
The Marcus equation is equivalent to the small polaron model in organic solids by reinterpreting 
the reorganization energy. \cite{Holstein_59,HOLSTEIN_59_2}
The reorganization energy in solution mainly originates from the coupling between 
the charge and solvent dipoles. 
In organic solids, it originates from the vibronic coupling in addition to the coupling between the charge and 
surrounding dipoles. 
Recently, the Marcus equation has been applied to study carrier transport in disordered molecular solids. 
\cite{Soos_00,Seki_01,Verbeek_92,Baranovskii14}
In 1 D, 
$C_d=3/4$ was obtained by analytical exact calculation based on the mean first passage time 
using the Marcus equation and the Gaussian density of states. \cite{Seki_01}
This value is different from $C_d=1$ obtained using MA process. 
For higher dimension, 
the value of $C_d$ is still controversial. 
The obtained values vary between $1/8$ and $0.6$, and there are some reports 
that $C_d$ depends on the value of the reorganization energy. \cite{Cottaar_11,Fishchuk_03,Fishchuk_13,Radin_15,Baranovskii14}

In this manuscript, we study the effective mobility for 
2D square lattice (the coordination number z=4) and 3D cubic lattice (the coordination number z=6) 
using the Marcus equation and the Gaussian density of states. 
The effective mobility is approximately obtained 
by applying an effective medium approximation (EMA). 
In general, 
the self-consistency equation obtained by EMA is expressed as 
an integral equation. 
In this manuscript, the integral has been evaluated numerically, and also 
an analytical expression has been obtained by further approximating the integration. 
The result is expressed as a simple scaling form given by Eq. (\ref{eq:scaling1}). 
The validity of approximating integration is checked by comparison to the original self-consistency equation. 

The EMA employed in this study is known to give the exact results for 
the nearest neighbor hopping transport in periodic lattices both in the limit of $z=2$ (one dimensional periodic lattice) and 
$z \rightarrow \infty$. \cite{Haus_87,Kehr_96}
However, the EMA results are approximate for other values of the coordination number. 
To assess the quality of the EMA approximation, we have performed  kinetic Monte-Carlo simulations and compared the results with those obtained by EMA.

In Sec. II, we show EMA results. 
In Sec. III, 
the results of EMA are compared with those obtained by kinetic Monte-Carlo simulations. 
In Sec. IV, we discuss our results, and in Sec. V we apply them to analyze experimental data. 
The conclusion is given in Sec. VI. 

\section{Theory}
\label{sec:II}
When carrier transport occurs by incoherent 
hopping transitions of a small polaron between adjacent molecules,  
the transition rate from the site denoted by $i$ to that denoted by $j$ can be given by 
the Marcus equation, \cite{Marcus_56,Marcus_64}
\begin{align}
\Gamma_{ij} (\Delta E_i) = \frac{2\pi}{\hbar}\frac{J^2}{\sqrt{4\pi \lambda k_{\rm B} T}} 
\exp \left(- \frac{\left(\Delta E_i+\lambda\right)^2}{4\lambda k_{\rm B} T} \right), 
\label{eq:Marcus}
\end{align}
where $\Delta E_i=E_j-E_i$, $E_i$ and $E_j$ are the site energies,  
$\hbar$ is the Planck constant divided by $2\pi$,  
$J$ is the transfer integral, and $\lambda$ is the reorganization energy.
In solid phases, the reorganization can be governed by both 
vibronic coupling \cite{Levich_59} and the dielectric relaxation of surroundings \cite{Marcus_56,Marcus_64}. 
For many molecular solids, 
the value of the reorganization energy can be    
$\lambda \sim 3-15 k_{\rm B} T$. \cite{Bredas_04}
The density of states of $E_i$ is assumed to obey the Gaussian distribution, 
\begin{align}
g(E_i)=\frac{1}{\sqrt{2\pi \sigma^2}} \exp \left(-\frac{E_i^2}{2 \sigma^2}\right).   
\label{eq:siteenergy}
\end{align}
The mean energy $\langle E_i \rangle$ can in principle be set to an arbitrary value since 
the Marcus equation is expressed by the site energy difference.  
Here, we set $\langle E_i \rangle=0$. 

For  the Gaussian density of states, the mean square displacement of a particle is known to be proportional to time, except for a certain non-stationary period. Such a behavior is confirmed by our simulations, as will be described later. If another form of the density of states is considered, described by a heavy-tailed exponential function, 
the transient non-stationary period will be prolonged. 
\cite{Barkai_98,Harvey_91,Berlin_93}
During the non-stationary period, 
the mean square displacement is not proportional to time and 
the diffusion coefficient is no longer a constant. 
\cite{Barkai_98,Harvey_91,Berlin_93}
Here, we focus on the effect of random energies on the diffusion constant of normal diffusion and will not study the effect of a heavy tailed distribution leading to the anomalous diffusion. 

The transition rate  $\Gamma (0)$
in the absence of the site energy distribution is obtained as
\begin{align}
\Gamma (0) = \frac{2\pi}{\hbar}\frac{J^2}{\sqrt{4\pi \lambda k_{\rm B} T}} \exp \left(- \frac{\lambda}{4 k_{\rm B} T} \right),  
\label{eq:Gamma0}
\end{align}
where the activation energy of hopping is given by $\lambda/4$. 

In the below, we consider the mobility of a single carrier on a hypercubic lattice. 
The coordination number of the lattice is denoted by $z$. 
We have $z=2d$ for a $d$-dimensional hypercubic lattice. 
On each site, a random site energy is assigned and the distribution is given by Eq. (\ref{eq:siteenergy}). 
Because the Marcus equation depends on the site energy, 
the carrier mobility differs for each realization of random site energy.
The effective mobility can be defined as its ensemble average. 
In EMA, 
the self-consistency condition is imposed to obtain the effective transition rate. 

The relation between the diffusion constant and the transition rate 
in the absence of the site energy distribution is given by $D_0=a^2 \Gamma(0)$, 
where $a$ is the lattice constant. 
The effective diffusion constant can be expressed using the effective transition rate by 
$D_{\rm eff}=a^2 \Gamma_{\rm eff}$. 
The ratio becomes  
\begin{align}
\frac{D_{\rm eff}}{D_0}=\frac{\Gamma_{\rm eff}}{\Gamma(0)}. 
\label{eq:ratioD}
\end{align}

The mobility satisfies the Einstein relation in the absence of the site energy distribution in the zero field limit 
$D_0=\mu_0 k_{\rm B} T/e$. 
The effective mobility also satisfies the Einstein relation 
$D_{\rm eff}=\mu_{\rm eff} k_{\rm B} T/e$ 
in 1 dimension in the zero field limit. \cite{Derrida}
In the higher dimension, 
the Einstein relation is numerically confirmed under certain conditions in the dilute limit. \cite{Haus_87}
Since we are interested in zero field mobility and the Einstein relation holds under linear response, 
we can safely assume  
\begin{align}
\frac{\mu_{\rm eff}}{\mu_0}=\frac{\Gamma_{\rm eff}}{\Gamma(0)}  
\label{eq:ratiomu}
\end{align}
and calculate $\Gamma_{\rm eff}/\Gamma(0)$ to obtain the mobility ratio given by $\mu_{\rm eff}/\mu_0$, 
where $\Gamma(0)$ is given by Eq. (\ref{eq:Gamma0}). 

The Einstein relation results from a linear response theory 
for stationary processes, 
so it is applicable when the external electric field is sufficiently small. \cite{Toda_92}
The condition of the weak field depends on the energetic disorder. 
\cite{Richert_89,Bouchaud_89,Derrida}
A stronger electric field dependence was found for the effective diffusion constant compared to that of the effective mobility. \cite{Richert_89,Bouchaud_89}
It should also be noted that 
the Einstein relation does not hold at short times before the process becomes stationary. 
This period again depends on the degree of energetic disorder. 
\cite{Schirmacher,BERLIN_96,Berlin_93,Berlin_99,Barkai_98,Harvey_91}
We confirm the stationarity of the processes considered in this study by analyzing the simulation results obtained over wide ranges of time.

In the simplest EMA, we consider random energy for two neighboring sites and 
ensemble average of a single transition rate connecting these sites is calculated while  
other transitions are expressed by an effective transition rate. 
The self-consistency condition is 
that the average over the different realizations of the random energy of two neighboring sites will reproduce the effective transition rate. 
When a single transition rate between a pair of neighboring sites is allowed to fluctuate and
these sites are embedded in the effective medium,   
these two random sites should be statistically equivalent. 
As shown in Appendix A, 
the EMA can be simplified, 
if the rate is symmetrized. \cite{Haus_87,Kehr_96}
The symmetrized rate in view of the detailed balance can be given by 
\begin{align}
\Gamma^{\rm sym} = \rho_i^{\rm(eq)} \Gamma_{ij},
\label{eq:symmetricrates}
\end{align}
where we abbreviated $\Gamma_{ij}^{\rm sym}$ by $\Gamma^{\rm sym}$. 
The abbreviation will not introduce confusion since only a single transition rate fluctuates. 
The equilibrium occupation probability at site $i$ denoted by $\rho_i^{\rm(eq)}$ can be expressed as
\begin{align}
\rho_i^{\rm(eq)}=\frac{\exp[-E_i/(k_{\rm B} T)]}{\langle \exp[- E_i/(k_{\rm B} T)] \rangle}=
\exp\left[-\frac{E_i}{k_{\rm B} T}
-\frac{1}{2}\left(\frac{\sigma}{k_{\rm B} T}\right)^2
\right]. 
\label{eq:rhoeq}
\end{align}
By using the Marcus hopping rate, 
$\Gamma^{\rm sym}$ can be explicitly written as  
\begin{align}
\Gamma^{\rm sym}=\frac{2\pi}{\hbar}\frac{J^2}{\sqrt{4\pi \lambda k_{\rm B} T}} 
\exp \left(- \frac{\left(E_j-E_i\right)^2}{4\lambda k_{\rm B} T}- \frac{E_j+E_i}{2k_{\rm B} T} -\frac{\lambda}{4k_{\rm B} T}-
\frac{\sigma^2}{2(k_{\rm B} T)^2}\right).
\label{eq:symMarcus}
\end{align}

The self-consistency condition is given by (see Appendix A)\cite{Kirkpatrick_73,Haus_87,Kehr_96}
\begin{align}
\left\langle \frac{\Gamma_{\rm eff}-\Gamma^{\rm sym}}{(z/2-1) \Gamma_{\rm eff}+ \Gamma^{\rm sym}} \right\rangle=0, 
\label{eq:selfconsistent}
\end{align}
where $z$ is the coordination number, $\Gamma_{\rm eff}$ denotes the effective mobility and 
$\langle \cdots \rangle$ denotes the ensemble average expressed by  
\begin{align}
\langle \cdots \rangle = \int_{-\infty}^\infty dE_i \int_{-\infty}^\infty dE_j 
\frac{1}{2\pi \sigma^2} 
\exp \left( -\frac{E_i^2+E_j^2}{2 \sigma^2} \right) \cdots.  
\label{eq:av}
\end{align}

When $z=2$ (1D), Eq. (\ref{eq:selfconsistent}) reduces to \cite{Kehr_96}
\begin{align}
\frac{1}{\Gamma_{\rm eff}}=  \left\langle \frac{1}{\Gamma^{\rm sym}} \right\rangle. 
\end{align}
The result is the same as the exact one obtained using the mean first passage time expressed as, \cite{Seki_01}
 \begin{align}
\frac{\Gamma_{\rm eff}}{\Gamma(0)} = \exp\left[-\frac{3}{4} \left(\frac{\sigma }{k_{\rm B}T}\right)^2 \right],
\label{eq:scaling1_1d}
\end{align} 
where $\Gamma(0)$ is given by Eq. (\ref{eq:Gamma0}) and is proportional to 
$\exp[-\lambda/(4 k_{\rm B} T)]/\sqrt{\lambda k_{\rm B} T}$. 

To solve analytically the self-consistency condition for $z>2$, we rewrite the self-consistency condition as
\begin{align}
\frac{1}{z/2-1}\left\langle 1- \frac{d \Gamma^{\rm sym}}{(z/2-1) \Gamma_{\rm eff}+ \Gamma^{\rm sym}}
\right\rangle=0. 
\label{eq:selfcc}
\end{align}
By rearrangement, we finally obtain
\begin{align}
\frac{2}{z}  = \left\langle \frac{1}{1+(z/2-1)\Gamma_{\rm eff}/\Gamma^{\rm sym} } \right\rangle.
\label{eq:selfcons_basic}
\end{align}
Here, we note that the factor $1/[1+(z/2-1)\Gamma_{\rm eff}/\Gamma^{\rm sym}]$ resembles Fermi-Dirac distribution function, 
which we will study closely.  

In order to see the pure influence of the random site energy, 
we introduce a normalized transition rate defined by,
\begin{align}
\Gamma_{\rm r} (\Delta E_i)=\frac{\Gamma_{ij} (\Delta E_i)}{\Gamma (0)} =
\exp \left(- \frac{(\Delta E_i)^2}{4\lambda k_{\rm B} T}- \frac{\Delta E_i}{2k_{\rm B} T} \right).
\label{eq:Gammar}
\end{align}
We can express $\Gamma^{\rm sym}/\Gamma_{\rm eff}$ as
\begin{align}
\frac{\Gamma^{\rm sym}}{\Gamma_{\rm eff}}=\frac{\Gamma_{\rm r} (\Delta E_i)\exp[-E_i/(k_{\rm B} T)]}{G_{\rm eff}}, 
\label{eq:convert}
\end{align}
where we defined 
\begin{align}
G_{\rm eff}=\Gamma_{\rm eff} \langle \exp[-E_i/(k_{\rm B} T)] \rangle/\Gamma (0). 
\label{eq:Geff}
\end{align}
Equation (\ref{eq:selfcons_basic}) can be reexpressed as
\begin{align}
\frac{2}{z}  = \left\langle \frac{1}{1+\exp \left[\left(E_i-\eta(\Delta E_i) \right)/(k_{\rm B} T)\right] } \right\rangle, 
\label{eq:selfcons_basic1}
\end{align}
where $\eta(\Delta E_i)$ is defined by, 
\begin{align}
\eta(\Delta E_i)&= - k_{\rm B} T \ln \left[(z/2-1) G_{\rm eff}/\Gamma_{\rm r} (\Delta E_i)\right] 
\label{eq:chemp}\\
&=- k_{\rm B} T \ln \left[\left(\frac{z}{2}-1\right) G_{\rm eff}\right] -  \frac{(\Delta E_i)^2}{4\lambda}- \frac{\Delta E_i}{2}.
\label{eq:chemp1}
\end{align}
Equation (\ref{eq:selfcons_basic1}) can be further rearranged into
\begin{align}
\frac{2}{z}  = \left\langle \frac{1}{1+\exp 
\left[\left(E_i+\frac{\Delta E_i}{2}+\frac{(\Delta E_i)^2}{4\lambda}-\eta_0 \right)/(k_{\rm B} T)\right] } \right\rangle, 
\label{eq:selfcons_basic1_r}
\end{align}
where $\eta_0$ is defined by 
\begin{align}
\eta_0=- k_{\rm B} T \ln \left[\left(\frac{z}{2}-1\right) G_{\rm eff}\right] .
\label{eq:chemp1}
\end{align}
The quantity inside $\left\langle \cdots \right\rangle$ in Eq. (\ref{eq:selfcons_basic1_r}) can be approximated as $1$ 
when $E_i+\Delta E_i/2+(\Delta E_i)^2/(4\lambda)$ is smaller than $\eta_0$
and decreases to zero as the value of $E_i+\Delta E_i/2+(\Delta E_i)^2/(4\lambda)$ increases over that of $\eta_0$. 
In this sense, 
$\eta_0$ plays a similar role to the chemical potential in Fermi-Dirac distribution function. 
Note that 
the value of $\eta_0$ can be determined 
for a given value of $G_{\rm eff}$ and $z$. 
The percolation path for the given value of $\eta_0$
consists of 
random energies satisfying 
$E_i+\Delta E_i/2+(\Delta E_i)^2/(4\lambda)\leq\eta_0 $. 
The interpretation of EMA results in terms of a percolation path was 
previously discussed for the transition rates used to study ion transport. 
\cite{Schirmacher}
We also note that Eq. (\ref{eq:av}) can be rewritten as 
\begin{align}
\langle \cdots \rangle = \int_{-\infty}^\infty d\Delta E_i \int_{-\infty}^\infty dE_i 
\frac{1}{2\pi \sigma^2} 
\exp \left( -\frac{(E_i+\Delta E_i/2)^2}{\sigma^2} - \frac{\Delta E_i^2}{4 \sigma^2} \right) \cdots.  
\label{eq:av1}
\end{align}
The average with respect to $E_i$ is given by a Gaussian function whose maximum is at $-\Delta E_i/2$.

We need different approximation to evaluate the integration with respect to $E_i$ depending on 
the value of the maximum given by $-\Delta E_i/2$ and $\eta(\Delta E_i)$. 
The condition $\eta(\Delta E_i)<-\Delta E_i/2$ can be expressed as 
\begin{align}
\ln \left[ \frac{\Gamma_{\rm r} (\Delta E_i)}{(z/2-1)G_{\rm eff}}
\right] < -\frac{\Delta E_i}{2 k_{\rm B} T} . 
\label{eq:cond1}
\end{align}
For the Marcus rate equation, 
Eq. (\ref{eq:cond1}) can be expressed 
using Eq. (\ref{eq:Gammar})  as
\begin{align}
\exp \left(- \frac{(\Delta E_i)^2}{4\lambda k_{\rm B} T} \right) < \left(\frac{z}{2}-1\right)G_{\rm eff} . 
\label{eq:cond2_1}
\end{align}
We note that Eq. (\ref{eq:cond2_1}) holds for $z\geq 4$ at least when 
$\sigma$ is small so that $G_{\rm eff} \sim 1$.
Therefore,  $\eta(\Delta E_i)<-\Delta E_i/2$ is the appropriate condition for $z\geq 4$. 

\begin{figure}
\includegraphics[width=1\columnwidth]{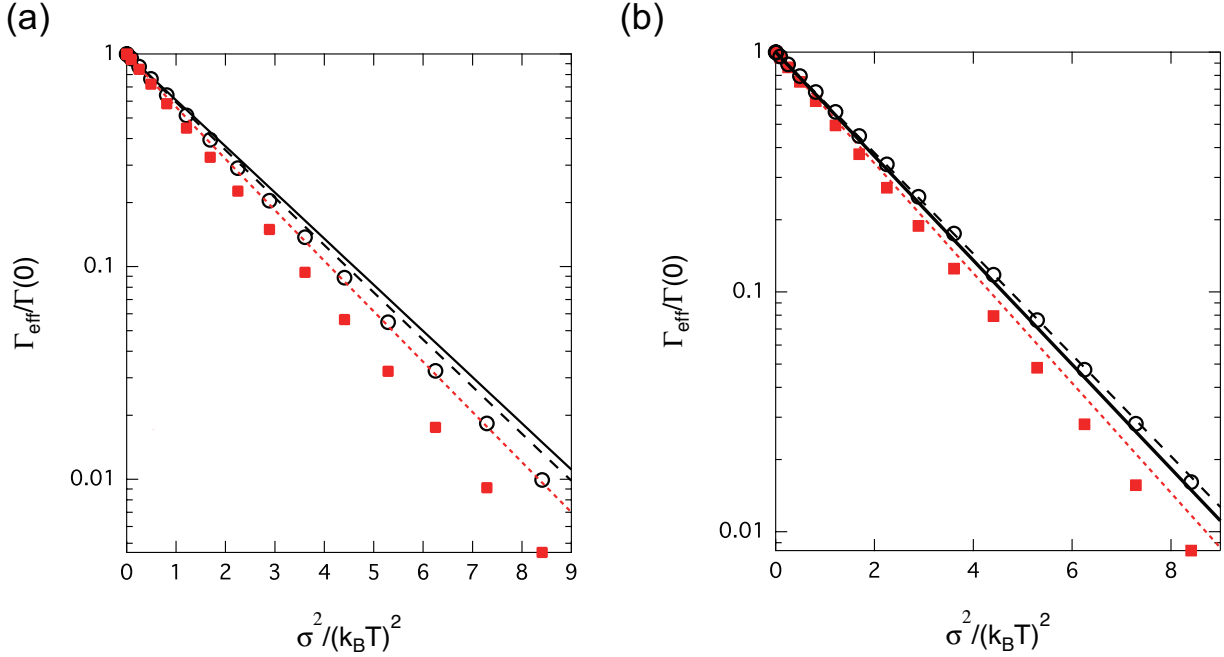}
\caption{$\Gamma_{\rm eff} /\Gamma (0)$ plotted as a function of $\sigma^2/(k_{\rm B} T)^2$. 
(a) 2D ($z=4$) and (b) 3D ($z=6$). 
The solid lines represent the scaling relation with $C_d=1/2$ obtained using EMA and given by Eq. (\ref{eq:selfcons_basic3}). 
Black circles and long dashed line indicate semi-analytical EMA results obtained for $\lambda/(k_{\rm B} T)=10$ 
by numerically evaluating Eq. (\ref{eq:selfconsistent}) 
(double integration) and Eq. (\ref{eq:selfcons_basic2}) (single integration), respectively. 
Red squares and red short dashed line indicate analogous semi-analytical EMA results obtained 
for $\lambda/(k_{\rm B} T)=3$. 
}
\label{fig:1}
\end{figure}

When $\eta(\Delta E_i)<-\Delta E_i/2$, we can employ the saddle point method to reduce the double integration in Eq. (\ref{eq:selfcons_basic1}) to single integration 
\begin{align}
\frac{2}{z}  = \frac{1}{2\sqrt{\pi \sigma^2}} 
\int_{-\infty}^\infty d\Delta E_i
\frac{\exp \left[ - \Delta E_i^2/\left(4 \sigma^2\right)\right]}
{1+(z/2-1) G_{\rm eff}\exp \left[-\Delta E_i/\left(2k_{\rm B} T\right)\right]/\Gamma_{\rm r} (-\Delta E_i/2) } .  
\label{eq:selfcons_basic2}
\end{align}
We numerically confirm the solution of Eq. (\ref{eq:selfcons_basic2}) 
by comparison with that of the original self-consistency equation given by Eq. (\ref{eq:selfconsistent}) in Fig. \ref{fig:1}. 
When $\lambda/(k_{\rm B} T)=10$, we find quite good agreement. 
When $\lambda/(k_{\rm B} T)=3$, some deviation is observed. 

Furthermore, when $\sigma$ is small we can again employ the saddle point method and obtain 
$2/z \sim 1/[1+(z/2-1) G_{\rm eff}]$, where we have used 
$\Gamma_{\rm r} (0)=1$. By introducing the definition of $G_{\rm eff}$ given by Eq. (\ref{eq:Geff}), we obtain a scaling relation, 
\begin{align}
\Gamma_{\rm eff} /\Gamma (0) \approx 1/\langle \exp[-\beta E_i/(k_{\rm B} T)] \rangle
= \exp\left[-\frac{1}{2}\left(\frac{\sigma}{k_{\rm B} T}\right)^2\right],  
\label{eq:selfcons_basic3}
\end{align}
where the transition rate in the absence of disorder $\Gamma(0)$ is given by Eq. (\ref{eq:Gamma0}).
As shown in Fig. \ref{fig:1}, the simple scaling relation of Eq. (\ref{eq:selfcons_basic3}) gives very close result to that obtained from 
the original self-consistency equation Eq. (\ref{eq:selfconsistent}) when $\lambda/(k_{\rm B} T)=10$. 
When $\lambda/(k_{\rm B} T)=3$, the degree of accuracy of the scaling relation is reduced. 
In the following, we study the validity of the scaling relation by using kinetic Monte-Carlo simulations. 

\section{Simulation results}
\label{sec:III}

\begin{figure}
\includegraphics[width=1\columnwidth]{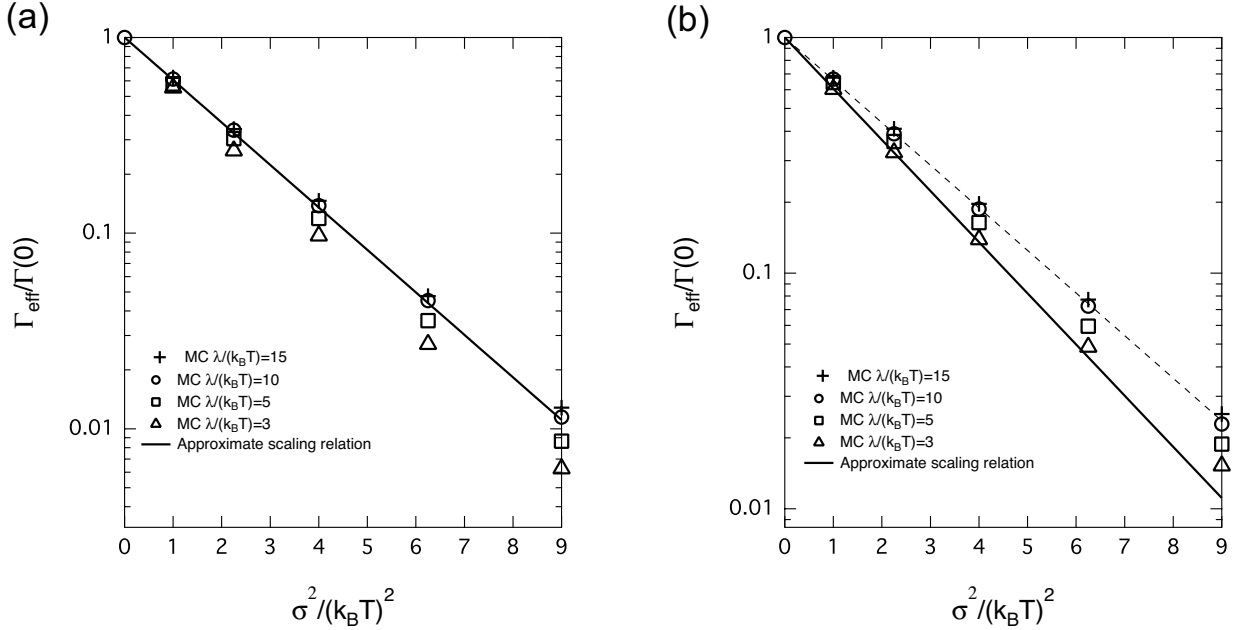}
\caption{$\Gamma_{\rm eff} /\Gamma (0)$ plotted as a function of $\sigma^2/(k_{\rm B} T)^2$. 
(a) 2D ($z=4$) and (b) 3D ($z=6$). 
The line represents the scaling relation given by Eq. (\ref{eq:selfcons_basic3}) obtained using EMA.
The crosses, circles, squares, and triangles indicate the kinetic Monte-Carlo simulation results for 
$\lambda/(k_{\rm B} T)=15, 10, 5, 3$, respectively. 
The dashed line in (b) indicates the result of fitting to Eq. (\ref{eq:fitt}) when $\lambda/(k_{\rm B} T)=10$. 
$C_d=0.42$ is obtained from fitting.  
}
\label{fig:2}
\end{figure}

The simulation is carried out on a square lattice ($z=4$) or a cubic lattice ($z=6$), 
with the lattice constant being assumed as $a=1$. 
A particle is initially placed at site (0,0) or (0,0,0), respectively. 
The energy at this site ($E_i$) and the energies at all nearest neighbor sites ($E_j$, $j=1,2,\cdots, z$) 
are sampled from the normal distribution $N(0,\sigma)$. 
The transition rates to the nearest neighbor sites, $\Gamma_{ij}$, are calculated from Eq. (\ref{eq:Marcus}), 
where the frequency factor $\nu_0=\left(2\pi/\hbar\right)J^2/\sqrt{4\pi \lambda k_{\rm B} T}$ is assumed equal to one. 
It is randomly decided to which of the nearest neighbor sites the particle will hop, with the probability of each hop being proportional to the corresponding transition rate $\Gamma_{ij}$. 
The time for the hop is sampled from an exponential distribution with the mean value 
$\tau=\left( \Gamma_{\rm tot}\right)^{-1}$, where $\Gamma_{\rm tot}=\sum_{j=1}^z \Gamma_{ij}$. 
The selected hop is now executed, and the procedure of sampling energies for new nearest neighbor sites (if not sampled before), 
calculating the transition rates, selecting the next hop, and so on, is repeated. 
The simulation run is carried out until the assumed total time $t_{\rm sim}$  is reached, 
and the squared distance of the particle from the origin $r^2(t_{\rm sim})$  is then recorded. 
The energies that are assigned to the lattice sites are kept in the memory for the whole duration of the simulation run. 
The simulation is repeated for $\sim 10^4$ independent runs to obtain the mean value $\langle r^2(t_{\rm sim}) \rangle$. 
The effective diffusion constant, relative to $D_0$, is then calculated as 
\begin{align}
\frac{D_{\rm eff}}{D_0}=\frac{\langle r^2(t_{\rm sim}) \rangle}{t_{\rm sim}  a^2 z \nu_0 \exp \left[-\lambda/(4 k_{\rm B} T) \right]}. 
\label{eq:sim1}
\end{align}
$D_{\rm eff}/D_0$ is essentially equivalent to $\Gamma_{\rm eff}/\Gamma(0)$ (cf. Eq. (\ref{eq:ratioD})). 
The simulation time $t_{\rm sim}$ has to be sufficiently long so that the long-time limit of Eq. (\ref{eq:sim1}) can be achieved. 
We analyzed the dependence of $D_{\rm eff}/D_0$  on $t_{\rm sim}$ for each set of the parameters, 
and found that it shows a decreasing trend at small values of $t_{\rm sim}$. 
For the final results presented in Fig. \ref{fig:2}, sufficiently long simulation times were chosen, for which this decreasing trend could no longer be observed.

The simulation results are compared with the  
EMA results in Fig. \ref{fig:2}. 
For 2D ($z=4$), 
the simulation results and that of the scaling relation given by Eq. (\ref{eq:selfcons_basic3}) coincide 
 for $\lambda/(k_{\rm B} T) \geq 10$.  
When the value of $\lambda/(k_{\rm B} T)$ is below $10$, 
the simulation results of $D_{\rm eff} /D (0)$ depend on $\lambda/(k_{\rm B} T)$ and 
are below 
the line drawn using Eq. (\ref{eq:selfcons_basic3}). 

For 3D ($z=6$), 
the results of kinetic Monte-Carlo simulations are independent of $\lambda/(k_{\rm B} T)$ when $\lambda/(k_{\rm B} T) \geq 10$.
Unlike in the case of 2D ($z=4$), 
the line drawn using Eq. (\ref{eq:selfcons_basic3}) is now below the simulation results. 
If we assume that the activation energy is not influenced by random energy and is expressed by $E_a=\lambda/4$, 
we obtain $C_d=0.42$ by fitting to 
\begin{align}
D_{\rm eff}/D_0=\exp\left[-C_d \sigma^2/(k_{\rm B} T)^2\right] 
\label{eq:fitt}
\end{align}
when $\lambda/(k_{\rm B} T) \geq 10$. 
This value is smaller than $C_d=1/2$ obtained from the scaling relation given by Eq. (\ref{eq:selfcons_basic3}). 
When $\lambda/(k_{\rm B} T)$ is below $10$, 
the simulation results of $\Gamma_{\rm eff} /\Gamma (0)$ depend on $\lambda/(k_{\rm B} T)$ and 
approach the line drawn using Eq. (\ref{eq:selfcons_basic3}) 
when the value of $\lambda/(k_{\rm B} T)$ decreases. 

The results of EMA show systematic deviation from the simulation results depending on $\lambda/(k_{\rm B} T)$ and the coordination number $z$, although the magnitude of this deviation is not large.
The deviation could originate from the use of the simplest version of EMA. 
In the simplest version of EMA, 
only a single transition rate is under the influence of random energy. 
The random energy in other sites are taken into account by the representative random transition rate in the effective medium.  
The accuracy of this approximation 
depends on the coordination number and the value of the reorganization energy as shown in Fig.  \ref{fig:2}.

\section{Discussion}

The effective mobility relative to $\mu_0$ is independent of $\lambda/(k_{\rm B} T)$ in 1D. \cite{Seki_01}
For higher dimensions ($z>2$), $\mu_{\rm eff}/\mu_0$ depends on $\lambda/(k_{\rm B} T)$ when $\lambda/(k_{\rm B} T)<10$.
When $\lambda/(k_{\rm B} T)\geq 10$, 
Eq. (\ref{eq:fitt}) with $C_d=1/2$ reproduces the simulation results of 2D ($z=4$) and 
$C_d=0.42$ is obtained from fitting to the simulation results for 3D ($z=6$). 

So far, various values of $C_d$ were reported for the Marcus transition rate by 
assuming Eq. (\ref{eq:scaling1}) in 3D.  
Using a different form of EMA self-consistency equation, 
Fishchuk {\it et al.} obtained $C_d=1/8$ 
for 3D when $\lambda/2>\sigma$. \cite{Fishchuk_03}
Later, it was suggested that $C_d$ value varies between 
$0.25-0.44$ depending on $\lambda/\sigma$. \cite{Fishchuk_13}
Recently,  a scaling form of Eq. (\ref{eq:scaling1}) with $C_d=1/2$ was proposed 
using a concept of fat percolation. \cite{Cottaar_11}
In the fat percolation theory, 
$E_a$ may contain contribution from random site energy and can be different from $\lambda/4$. 
The results of fat percolation theory were compared to the numerical results obtained using 
the master equation method. \cite{Cottaar_11}
The obtained numerical  values of $C_d$ were in the range between $0.69-0.44$ for simple cubic lattice
by regarding $E_a$ as a free parameter for fitting. \cite{Cottaar_11}
$C_d$ values determined from fitting can be influenced by $E_a$ values. 
We share a conclusion of scaling with $C_d=1/2$ for simple cubic lattice obtained by the fat percolation theory. 
There could be subtle issues regarding how $E_a=\lambda/4$ and $C_d=1/2$ 
should be corrected for the simple cubic lattice, 
where  $16 \%$ smaller value of $C_d$ is obtained by fitting to the results of kinetic Monte-Carlo simulations 
using $E_a=\lambda/4$ for $\lambda/(k_{\rm B} T)\geq 10$.  
In this study, 
an analytical expression was approximately derived from the self-consistency equation of EMA. 
In the fat percolation theory, 
an additional dependence of $E_a$ on $\sigma$ can be considered. \cite{Cottaar_11}
The correction term is too small compared to the accuracy of EMA used in this study. 
For simplicity, we put  $E_a=\lambda/4$ to determine $C_d$ using kinetic Monte-Carlo simulations.
More elaborate theories are required to study such deviations.

Very recently, $E_a=\lambda/4$ and $C_d=1/4$ have been suggested as the upper bound 
using the generalized effective medium theory. 
\cite{Radin_15}
We can obtain $E_a=\lambda/4$ and $C_d=1/4$ by taking  
$z \rightarrow \infty$ limit in EMA.  (see Appendix B)
For simple cubic lattice we have $z=6$. 
The value of $z=6$ is too small to regard it as $z \rightarrow \infty$. 
As a result, the result of EMA for $z=6$ is very different from that 
obtained by taking the limit of $z \rightarrow \infty$. 

We focused on the effective mobility when the carrier concentration is low. 
At high carrier concentration, 
one should note that 
carrier transitions are not allowed if the target sites are occupied. 
When the effective mobility is obtained under the steady state at high carrier concentration, 
low energy states are filled. 
Since the part of density of states below a certain energy is mainly occupied, 
the unoccupied density of states differs from the density of states that includes occupied states.
The carrier mobility increases by increasing the carrier concentration when the filling effect sets in.  \cite{Cottaar_11,Fishchuk_13,Lu_15}
Recently, it was under debate whether 
$C_d$ depends on the ratio between $\lambda$ and $\sigma$ at high carrier concentration.  \cite{Cottaar_11,Fishchuk_13,Lu_15}
In Ref. \onlinecite{Fishchuk_13}, the dependence of $C_d$ on the ratio between $\lambda$ and $\sigma$ was obtained 
by Monte-Carlo simulations and an effective medium theory with an averaging method different from 
that employed here. 
At sufficiently low carrier concentration, their results and ours should coincide. 
Unfortunately, since the concentration dependence of $C_d$ is unclear, 
the results of Ref. \onlinecite{Fishchuk_13} cannot be directly compared with ours.

\section{ANALYSIS OF EXPERIMENTAL DATA}
In this Section, 
the theoretical results obtained in the present study are applied to analyze the experimental data. 
We show two examples of such an analysis, 
in which we interpret the results of hole mobility measured in 2D and 3D systems.
We assume that the effective mobility can be expressed as 
\begin{align}
\mu_{\rm eff}=\frac{C_{\mu}}{\lambda^{1/2} \left(k_{\rm B} T\right)^{3/2}}\exp\left[-\frac{\lambda}{4k_{\rm B}T}-
C_d \left(\frac{\sigma}{k_{\rm B} T}\right)^2
\right], 
\label{eq:conclusion1}
\end{align}
where $C_{\mu}$ is a constant independent of $T$, $\lambda$ and $\sigma$. 
For the analysis of the 2D system, 
we use $C_d=0.5$, as obtained from both the EMA and Monte Carlo simulations at 
$\lambda \geq 10 k_{\rm B} T$. For the 3D system, we use 
$C_d=0.42$ obtained from the simulations when $\lambda \geq 10 k_{\rm B} T$.

Using the experimental data, 
we determine the values of the disorder parameter $\sigma$, 
and compare them with those obtained by the conventional method, where the Miller-Abrahams (MA) rate is used to describe the charge carrier transitions instead of the Marcus reaction rate. 
The MA rate is expressed as 
$\Gamma_{ij} (\Delta E_i) = \Gamma_0$ for $\Delta E_i\leq 0$ and 
$
\Gamma_{ij} (\Delta E_i) = \Gamma_0
\exp \left[- \Delta E_i/(k_{\rm B} T) \right] 
$ for $\Delta E_i> 0$, 
where $\Gamma_0$ is a constant independent of $T$ and $\Delta E_i$.
As shown by Monte Carlo simulations, when the MA rate is used to model the hopping transitions, 
the effective mobility for the cubic lattice is well described by
\begin{align}
\mu_{\rm eff}=C_{\mu}' \exp\left[-
C_{\rm MA} \left(\frac{\sigma_{\rm MA}}{k_{\rm B} T}\right)^2
\right], 
\label{eq:fitting2}
\end{align}
where $C_{MA}=0.44$. 
It should be noted that the activation energy 
$E_a$ does not appear in Eq. (\ref{eq:fitting2}). 
The charge carrier transport was interpreted in this case as 
an exclusively disorder-controlled ($E_a=0$) process. 
In the present study, we obtained $C_d=0.42$ with $E_a=\lambda/4$ and an additional algebraic $T$-dependence 
under the condition of  $\lambda \geq 10 k_{\rm B} T$. 

\begin{table}
\caption{\label{tab:table1}
Reorganization energy and disorder parameters obtained from temperature dependence of mobilities 
reported in Ref. \onlinecite{Hoffmann_13}. 
}
\begin{ruledtabular}
\begin{tabular}{lcccc}
copolymer\footnote{Ref. \onlinecite{Hoffmann_13}.}&
$\lambda$ [eV]\footnotemark[1]&
$\sigma$ [eV]\footnote{The values obtained using Eq. (\ref{eq:conclusion1}).}&
$\sigma_{\rm MA}$ [eV]\footnote{The values obtained using Eq. (\ref{eq:fitting2}).}& 
$\sigma/\sigma_{\rm MA}$ [\%]\\
\hline
1 & 0.3 & 0.095&0.109&87\\
3& 0.3 & 0.098&0.102&96\\
7 &0.2 & 0.074&0.089&83\\
9 & 0.3 & 0.065&0.091&71\\
\end{tabular}
\end{ruledtabular}
\end{table}
\begin{figure}
\includegraphics[width=0.5\columnwidth]{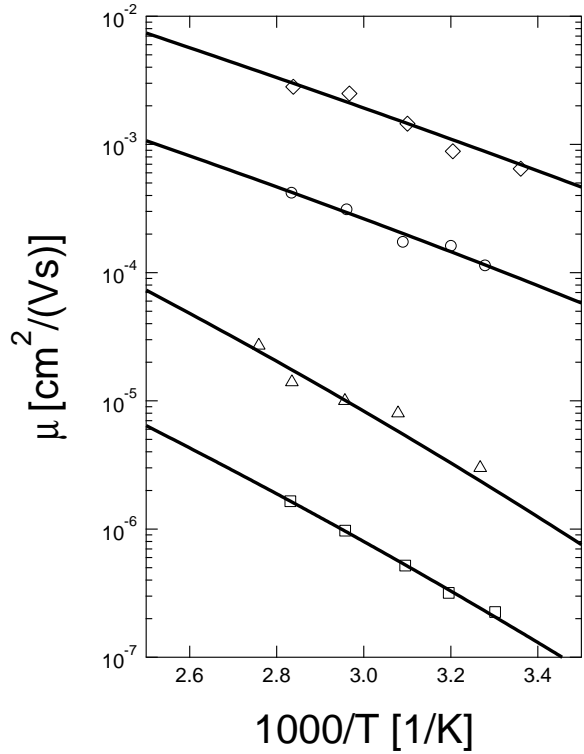}
\caption{Mobility plotted as a function of $1/T$ [1/K]. 
Squares, triangles, circles, and diamonds
indicate 
the experimental data obtained in \mbox{Ref. \onlinecite{Hoffmann_13}} 
for compounds 1, 3, 7, and 9, respectively.
The lines represent the results of fitting using 
Eq. (\ref{eq:conclusion1}) with $C_d=0.42$ and the values of $\lambda$ shown in Table \ref{tab:table1}. 
}
\label{fig:3}
\end{figure}
In recent experiments, both the reorganization energy and the effective hole mobility 
were measured in conjugated copolymers. \cite{Hoffmann_13}
Hole transport in conjugated copolymers can be regarded as random walks in 3D systems. 
For all copolymers, the values of reorganization energy were estimated in the range of $0.2 \sim 0.3$ eV as summarized in 
Table \ref{tab:table1}. 
These values approximately satisfy $\lambda \geq 10 k_{\rm B} T$. 
Therefore, Eq. (\ref{eq:conclusion1}) with $C_d=0.42$ is applicable. 
In Ref. \onlinecite{Hoffmann_13}, the experimental data were interpreted by assuming 
either exclusively polaronic ($\sigma=0$, $E_a\neq0$) or exclusively disorder-controlled ($E_a=0$) transport for the holes. 
It could be more natural to assume that the hole transport is both affected by disorder of the medium ($\sigma\neq0$) 
and displays a non-zero activation energy that originates from the reorganization energy. 
The latter was optically measured in Ref. \onlinecite{Hoffmann_13}, 
separately from the time-of-flight experiments performed to determine the effective mobility. 

We analyze 
4 types of conjugated alternating phenanthrene
indenofluorene copolymers denoted by 1,3,7, and 9 in Ref. \onlinecite{Hoffmann_13}. 
The reorganization energy obtained 
from an analysis of fluorescence spectra is given by 
$\lambda=0.3$  eV for copolymer 1,3,9 and $\lambda=0.2$  eV for copolymer 7. 
We fit Eq. (\ref{eq:conclusion1}) to the experimental data, 
as illustrated in Fig. 3, and determine the values of $\sigma$, 
which are listed in Table \ref{tab:table1} together with the values of $\sigma_{MA}$ reported in Ref. \onlinecite{Hoffmann_13}. 
The values of $\sigma$ are 
$4\sim29$ \% smaller than $\sigma_{MA}$.
These results indicate that when the reorganization energy  
is ignored, the disorder parameter $\sigma$ can be significantly overestimated. 
Regarding the question of whether the hole transport is polaronic or disorder-controlled, 
we note that the determined values of $\sigma$ and the thermal activation energy of polaron transport given by $E_a=\lambda/4$ 
are comparable. 
In this sense, both the reorganization energy and the energetic disorder affect the effective mobility. 

\begin{figure}
\includegraphics[width=0.5\columnwidth]{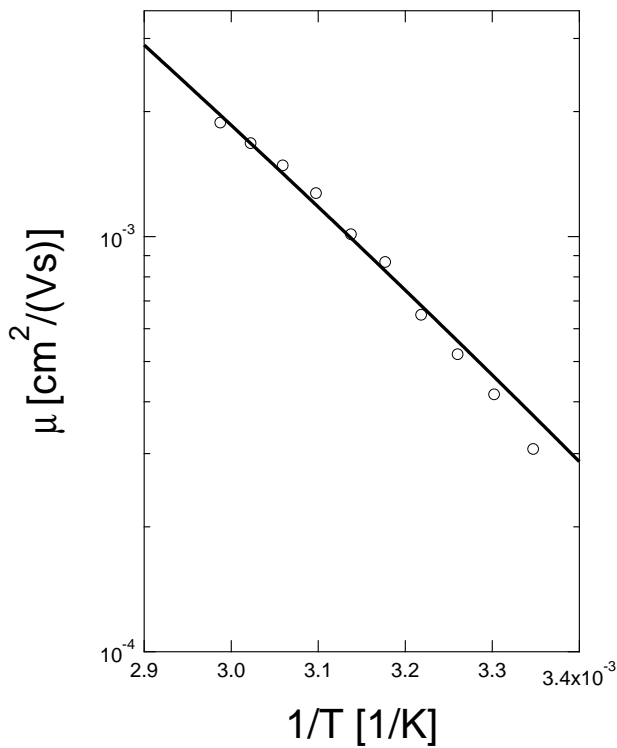}
\caption{2D hole mobility in SmE phase of a liquid crystal plotted as a function of $1/T$ [1/K]. 
The circles indicate experimental data taken from 
Ref. \onlinecite{Ohno_03}. 
The solid line is the result of fitting using Eq. (\ref{eq:conclusion1}) with $C_d=1/2$ and $\lambda=0.3$ eV. 
}
\label{fig:4}
\end{figure}

As an example of 2D charge carrier transport, 
we consider the hole transport in smectic liquid crystals. 
Smectic liquid crystals form layered structures and holes are expected to move within 
a layer. 
We analyze the temperature dependence of the hole mobility 
in 6O-BP-6 2D smectic mesophases of biphenyls 
reported in Ref. \onlinecite{Ohno_03}. 
In the temperature range shown in Fig. \ref{fig:4}, 
the liquid crystal is in SmE phase, where 
molecules form a rectangular lattice in each layer. 
For reorganization energy, 
we assume $\lambda=0.3$ eV, a typical value for organic molecules.  
This value satisfies $\lambda \geq 10 k_{\rm B} T$  
so we apply Eq. (\ref{eq:conclusion1}) with $C_d=1/2$ obtained for 2D carrier transport. 
By analyzing the experimental data, we obtain 
 $\sigma=0.089$ eV, which is  
19\% smaller than $\sigma_{\rm MA}=0.11$ eV 
obtained in Ref. \onlinecite{Ohno_03}. 
Our value of $\sigma$ is close to the range $0.05-0.06$ eV, which is considered as a typical range of the disorder parameter that characterizes the hole transport in smectic liquid crystals. \cite{Ohno_03} 

\section{Conclusion}
\label{sec:VI}
Using an effective medium approximation (EMA),  we have analytically derived the scaling relation given by 
Eq. (\ref{eq:conclusion1}).
Equation (\ref{eq:conclusion1}) describes the effective charge carrier mobility  
when the elementary transition rate is given by the Marcus equation and the density of states is expressed by a Gaussian. 
We have also performed kinetic Monte-Carlo simulations for 2D ($z=4$ square lattice) 
and 3D ($z=6$  cubic lattice) to obtain the parameter 
$C_d$ by fitting.
Our results can be summarized as follows. 

Previously, $C_d=3/4$ was derived for 1D systems. \cite{Seki_01}
We have now obtained 
$C_d=1/2$ for 2D ($z=4$), and $C_d=0.42$ for 3D ($z=6$) 
when $\lambda/(k_{\rm B} T) \geq 10$. 
The last value was obtained by kinetic Monte-Carlo simulations 
and is somewhat lower than our analytical result ($C_d=1/2$) obtained for the 3D system. 
We note that the value of $C_d$ for 1D systems
is very different from those obtained for other lattices of higher dimensionality. \cite{Seki_16}
This result reflects the unique nature of the trajectories of mobile particles in one dimensional periodic lattices. 
In one dimension,  
if a transition to a new site does not occur because of a high barrier,  
the mobile particle jumps back to the previously occupied site, 
but it will 
finally succeed to pass the barrier after many trials and a long enough time. 
When the standard deviation $\sigma$ of the energetic disorder is increased in 1D, 
the growth of the mean square displacements will be suppressed by repeated trials to overcome the high barriers. 
On the contrary, 
transitions over high barriers will be avoided by changing the direction of the particle motion in 2D and 3D. 
The large difference between the $C_d$ values for 1D and those for 2D and 3D 
can probably be explained by the above considerations.

The kinetic Monte-Carlo simulations confirmed 
the value $C_d=1/2$ obtained from the EMA for the 2D system. 
On the other hand, we see a 16\% difference in $C_d$ between the theory and simulation in 3D. 
This difference 
could originate from adoption of the simplest EMA, where a single transition rate fluctuates in the effective medium. 
Although the effect of the coordination number can be partly taken into account by 
the representative random transition rate in the effective medium, 
the accuracy will decrease by going from 2D to 3D. 

The value of $C_d=0.42$ for 3D (cubic lattice) is close to $C_d=0.44$ of MA process. \cite{Bassler_93}
In 1D,  $C_d=0.75$ is obtained using the Marcus equation while $C_d=1$ is obtained for the MA process. \cite{Cordes_01,Seki_01}
These results indicate that the difference decreases by increasing the coordination number and suggest 
that the universal scaling relation of the form given by Eq. (\ref{eq:conclusion1}) for $z>2$ could be 
less sensitive to the types of elementary transition rates compared to that in 1D. 
Recently, a similar scaling relation was proposed for the MA process in a different context. \cite{Seki_Bagchi_2015,Seki_16}

There is a subtle issue about determination of  the value of $C_d$ for 2D and 3D systems. 
Previously, the value of $C_d$ of MA process was determined by assuming that the 
activation energy $E_a$ is zero because the 
activation energy associated with the reorganization energy is absent. 
Although the reorganization energy is absent, 
an activation energy induced by energetic disorder 
$E_a=[1-(1/\sqrt{2})] \sqrt{\pi} \sigma$ 
was recently derived by 
applying EMA using the MA process for 2D systems. \cite{Seki_16}
The disorder induced activation energy is important when $\sigma \leq k_{\rm B} T$. 
Further theoretical studies of this effect are required, especially for 3D systems.

It should also be noted that 
$\Gamma_{\rm eff} /\Gamma (0)$ is insensitive to the value of $\lambda/(k_{\rm B} T)$ irrespective of the values of the coordination number 
when  $\lambda/(k_{\rm B} T) \geq 10$. 
However, when $\lambda/(k_{\rm B} T) < 10$, 
$\Gamma_{\rm eff} /\Gamma (0)$ depends on the value of $\lambda/(k_{\rm B} T)$ both for 2D ($z=4$) and 3D ($z=6$). 
This dependence can be seen both in the simulation results and the results obtained by 
numerically evaluating the self-consistency equation of EMA. 
According to the Marcus rate expression given by Eq. (\ref{eq:Marcus}), 
the dependence of the transition rate on $\Delta E_i$ increases by decreasing the value of $\lambda$. 
As a result, the effective rate is more affected by the site energy distribution when $\lambda$ is small. 
The effective rate in the absence of the site energy distribution is given by $\exp[-\lambda/(4 k_{\rm B} T)]/\sqrt{\lambda T}$ but 
the $\lambda$-dependence may be modified under the strong influence of the site energy distribution when 
$\lambda/(k_{\rm B} T)$ is not sufficiently large. 
Interestingly, such an extra $\lambda$-dependence is absent in the exact result of 1D ($z=2$). \cite{Seki_01}
Again, the one dimensional result is different from those in higher dimensions.

We have obtained the effective mobility in the limit of low carrier density. 
At high carrier density, some parts of the density of states are occupied by carriers and 
the distribution of unoccupied states is thereby distorted. 
The trap filling effect can be important under device operating conditions. 
In Eq. (\ref{eq:scaling1}), $E_a$ and $C_d$ may depend on the concentration of carriers if 
carrier concentration is 
above a threshold value. \cite{Baranovskii14,Cottaar_12,Fishchuk_13}
It is important to note that 
the results in this manuscript are valid if the carrier concentration is below a certain threshold concentration.

We did not note any results for 2D ($z=4$) reported previously. 
Our result obtained for 2D may be useful in analyzing real charge carrier transport processes, beyond theoretical interests. 
In general, molecular solids can be highly anisotropic in structure. \cite{Jakobsson,Stehr_11}
The carrier transport can also be anisotropic reflecting the structure.

In this study, 
we used the Marcus equation assuming the classical high temperature limit of 
quantum transport between localized states. 
We assumed incoherent hopping of a polaron formed as a result of localization due to electron-phonon coupling in organic solids. 
In the studies of high charge mobility 
in  molecular crystals such as pentacene and rubrene, 
the assumption of a hopping transport between localized states might be inadequate. 
Recently, the influence of delocalized states and dynamic disorder on the effective mobility has been studied extensively.
\cite{Troisi_03,Troisi_11,Troisi_09}
At low temperatures, 
the band transport disturbed by phonon scattering contributes to the particle diffusion in addition to 
the phonon-assisted hopping.\cite{Grover71,Kitahara76,Troisi_11}
If the temperature is sufficiently low so that the wave functions are delocalized, 
both the localization and the intrinsic transfer rates 
depend on the 
inhomogeneous disorder, dimensionality, temperature and can be anisotropic. \cite{Moix_13,Lee_15,Chuang_16}
The effect of dimensionality on the temperature dependence of the effective mobility at low temperatures 
requires further theoretical investigation on the coherence dephasing. 

\acknowledgments

This work was supported by JSPS KAKENHI Grant Number 15K05406. 
One of us (M.W.) acknowledges support from the National Science Center of Poland (Grant No. DEC-2013/09/B/ST4/02956).

\newpage
\renewcommand{\theequation}{A.\arabic{equation}}  
\setcounter{equation}{0}  
\section*{Appendix A. Derivation of Eq. (\ref{eq:selfconsistent}) for a symmetrized rate}
When EMA is formulated using a symmetrized transition rate,\cite{Haus_87,Kehr_96} 
an additional approximation is introduced for the symmetrization. 

We denote the position on a hypercubic lattice by $\vec{r}_i$. 
Transition between neighboring sites can be designated by the displacement vector $\vec{\ell}_k$, where 
$k$ runs from $1$ to the coordination number $z$.
We consider random site energy on the origin denoted by $\vec{r}_0$ and a neighboring lattice site denoted by $\vec{r}_1=\vec{\ell}_1$. 
The transition between these sites are given by the Marcus equation Eq. (\ref{eq:Marcus}) and expressed by
 $\Gamma_{0,1}(\Delta E_0)$ with $\Delta E_0=E_{\vec{\ell}_1}-E_0$ for the transition from the origin and 
 $\Gamma_{1,0}(\Delta E_1)$ with $\Delta E_1=E_0-E_{\vec{\ell}_1}$ for the transition from $\vec{r}_1$.
We study the effective transition rate for the time evolution of the density  $\rho(\vec{r}_i,t)$ 
expressed by \cite{Haus_87,Kehr_96}
\begin{align}
\frac{\partial}{\partial t} \rho(\vec{r}_i,t)&=\int_0^t dt_1 \Gamma_{\rm eff}^{\rm unsym} (t-t_1)
\sum_{k=1}^z \rho(\vec{r}_i+\vec{\ell}_k,t_1)- z\int_0^t dt_1 \Gamma_{\rm eff}^{\rm unsym} (t-t_1)
\rho(\vec{r}_i,t_1)-
\nonumber \\
&
\left[\Gamma_{j,1-j}(\Delta E_j)-\int_0^t dt_1 \Gamma_{\rm eff}^{\rm unsym} (t-t_1)\rho(\vec{r}_j,t_1)
\right] (\delta_{i,j}-\delta_{i+j,1})(\delta_{i,0}-\delta_{i,1}). 
\label{eq:apA1}
\end{align}
The initial condition is given by $\rho(\vec{r}_i,0)=\delta_{i,0}$. 
$\Gamma_{\rm eff}^{\rm unsym} (t)$ indicates the effective transition rate for the original unsymmetrized rate.
By the Laplace transformation, we obtain  
\begin{align}
s \hat{\rho}(\vec{r}_i,s)-\delta_{i,0}&=\hat{\Gamma}_{\rm eff}^{\rm unsym} (s)
\sum_{k=1}^z \hat{\rho}(\vec{r}_i+\vec{\ell}_k,s)- z \hat{\Gamma}_{\rm eff}^{\rm unsym} (s)
\hat{\rho}(\vec{r}_i,s)-
\nonumber \\
&
\left[\Gamma_{j,1-j}(\Delta E_j)\hat{\rho}(\vec{r}_j,s)-\hat{\Gamma}_{\rm eff}^{\rm unsym} (s)\hat{\rho}(\vec{r}_j,s)
\right] (\delta_{i,j}-\delta_{i+j,1})(\delta_{i,0}-\delta_{i,1}),
\label{eq:apA2}
\end{align}
where $\hat{f} (s)$ denotes the Laplace transform of arbitrary function $f(t)$. 
In the above,  $\Gamma_{0,1}(\Delta E_0)$ and $\Gamma_{1,0}(\Delta E_1)$ are not equal. 
The calculation of the effective rate requires the inverse transformation of 2x2 matrix equation and the final expression is tedious. 

The simpler expression can be obtained by introducing a symmetrized rate. 
To formulate EMA by introducing a symmetrized rate, 
we define reduced density by 
\begin{align}
Q(\vec{r}_i,t)=\rho(\vec{r}_i,t)/\rho_i^{\rm(eq)}
\label{eq:apA3}
\end{align}
and note 
\begin{align}
\Gamma_{i,j}(\Delta E_i) \rho(\vec{r}_i,t)=\Gamma^{\rm sym} Q(\vec{r}_i,t),
\label{eq:apA4}
\end{align}
where $\Gamma^{\rm sym}$ is given by Eq. (\ref{eq:symmetricrates}). 
If we introduce 
$\Gamma_{\rm eff,i}^{\rm sym} (t)=\Gamma_{\rm eff}^{\rm unsym} (t) \rho_i^{\rm(eq)}$ 
according to Eq. (\ref{eq:apA3}), 
Eq. (\ref{eq:apA2}) can be rigorously rewritten using $\Gamma_{\rm eff,i}^{\rm sym} (t)$ but 
$\Gamma_{\rm eff,i}^{\rm sym} (t)$ is not homogeneous. 
Instead, 
we introduce 
$\Gamma_{\rm eff} (t) \approx \Gamma_{\rm eff}^{\rm unsym} (t) \langle \rho_i^{\rm(eq)} \rangle$. 
Since we have 
 $\langle \rho_i^{\rm(eq)} \rangle=1$, 
 we obtain 
 $\Gamma_{\rm eff}(t) \approx \Gamma_{\rm eff}^{\rm unsym}(t)$. 
Under the approximation, Eq. (\ref{eq:apA2}) can be expressed as 
\begin{multline}
s \hat{Q}(\vec{r}_i,s)-\delta_{i,0}
=\hat{\Gamma}_{\rm eff} (s)
\sum_{k=1}^z \hat{Q}(\vec{r}_i+\vec{\ell}_k,s)- z \hat{\Gamma}_{\rm eff} (s)
\hat{Q}(\vec{r}_i,s)-
\\
\left[\Gamma^{\rm sym}\hat{Q}(\vec{r}_j,s)-\hat{\Gamma}_{\rm eff} (s)\hat{Q}(\vec{r}_j,s)
\right] (\delta_{i,j}-\delta_{i+j,1})(\delta_{i,0}-\delta_{i,1})+
s \left( 1 - \rho_i^{\rm(eq)} \right) \hat{Q}(\vec{r}_j,s). 
\label{eq:apA5}
\end{multline}
The effective rate $\Gamma_{\rm eff}$ obtained from Eq. (\ref{eq:apA5}) can be regarded as 
the effective rate $\Gamma_{\rm eff}^{\rm unsym}$ of Eq. (\ref{eq:apA2}) approximately. 
The above equation has the common structure of the simplest EMA except the last term which vanished in the limit of $s\rightarrow 0$. 
Equation (\ref{eq:selfconsistent}) 
can be derived from Eq. (\ref{eq:apA5}) by applying usual procedure. \cite{Haus_87,Kehr_96}

\renewcommand{\theequation}{B.\arabic{equation}}  
\setcounter{equation}{0}  
\section*{Appendix B. Derivation of the upper limit}

We rewrite Eq. (\ref{eq:selfconsistent}) as, 
\begin{align}
\frac{1}{\Gamma_{\rm eff}}=  \left\langle \frac{z/2}{(z/2-1) \Gamma_{\rm eff}+ \Gamma^{\rm sym}} \right\rangle. 
\label{eq:ap1}
\end{align}
We use a systematic expansion expressed by
\begin{align}
\left(X+Y \right)^{-1}=X^{-1}-X^{-1}Y\left(X+Y \right)^{-1} > X^{-1}-X^{-1}Y X^{-1}, 
\label{eq:ap2} 
\end{align}
where $X$ and $Y$ are arbitrary function, and  
$Y>0$ is assumed. 
By applying the expansion to Eq. (\ref{eq:ap1}) by setting $X=(z/2-1) \Gamma_{\rm eff}$ 
and $Y=\Gamma^{\rm sym}$, 
we obtain, 
\begin{align}
\Gamma_{\rm eff} >\frac{z}{z-2} \frac{1}{\Gamma_{\rm eff}} -
\left(\frac{z}{z-2}\frac{\Gamma (0)}{\Gamma_{\rm eff}} \right)^2 \frac{2}{z} \left\langle \Gamma^{\rm sym}  \right\rangle + \cdots . 
\label{eq:ap3}
\end{align}
The expansion is better as the coordination number increases, $z \gg 1$.
By rearrangement, Eq. (\ref{eq:ap3}) can be expressed as 
\begin{align}
\Gamma_{\rm eff} <\frac{z}{z-2} \langle \Gamma^{\rm sym} \rangle . 
\label{eq:expsol}
\end{align}
By using the Marcus rate equation, we obtain 
 \begin{align}
\frac{\Gamma_{\rm eff}}{\Gamma(0)} < \left(\frac{z}{z-2}\right)\frac{\Gamma (0)}{1+\sigma^2/(\lambda k_{\rm B} T)}
\exp\left(-\frac{\sigma^2}{4 (k_{\rm B} T)^2} \right)  .    
\label{eq:expapprox}
\end{align}
The upper limit is close to that proposed recently using a different method. \cite{Radin_15}

%

\end{document}